# Temperature-dependent Pattern Formation in Drying Aqueous Droplets of Lysozyme

*Anusuya Pal[1], Amalesh Gope[2*], and Germano S. Iannacchione[1]*

[1]Order-Disorder Phenomena Laboratory, Department of Physics, Worcester Polytechnic Institute, Worcester, MA, 01609, USA
[2]Department of English, Tezpur University, Tezpur, Assam, 784028, India

*Email: amaleshtezu@gmail.com*

**Abstract**

Drying colloidal droplets have a wide range of applications from medical diagnostics to coatings for industries. This paper explores the effects of the substrate temperature (ranging from 25 to 55 °C) and various initial concentrations ($\emptyset$) of 1 to 20 wt% of lysozyme in an aqueous solution on its drying and final dried film state using bright-field optical microscopy. The $\emptyset$ is divided into three regimes, ultra-concentrated ($20 < \emptyset \leq 17$ wt%), concentrated ($17 < \emptyset \leq 9$ wt%) and diluted ($9 < \emptyset \leq 1$ wt%). During the drying process, initially, the fluid front moves slowly and linearly inward followed by a fast, non-linear movement in both dilute and concentrated regimes. Increasing $\emptyset$ in these regimes finds that this movement in the later non-linear region slows down as the front carries and deposits protein molecules until the supply in solution is exhausted. In the ultra-concentrated regime, the fluid front moves linearly throughout the drying process. The deposition of protein onto the surface by the fluid front creates the "coffee-ring" and increases with increasing $\emptyset$. A dimple is observed in a central mound-structure, which grow with increase $\emptyset$. As the temperature increases the drying rate increases, reducing the time for convective flow and the deposition of material at the fluid front. Interestingly, at (T, $\emptyset$) = (55 °C, 20 wt%), the droplet forms the thickest film with suppressed ring formation. The dimple diminishes in the ultra-concentrated regime, whereas it changes to an expanded spot in some samples of the diluted and concentrated regimes with elevated temperatures. Both initial concentration and substrate temperature leads to surface tension and temperature gradients across the droplet, affecting the final dried crack morphological patterns. This study provides insights into protein-protein and protein-substrate interactions driven by temperature and concentration for biomedical and biosensing applications.

## 1. Introduction

Drying colloidal droplets have a wide range of industrial applications such as coatings, inkjet printing, etc. [1]. Furthermore, the colloids with biological relevance (bio-colloids) have the potential to be used in medical diagnostics as the final dried patterns can be linked to the nature and state of the constituent particles [2]. However, the understanding of the drying of bio-colloids is challenging because of the presence of the multi-components, their self-interacting interactions, and the non-equilibrium characteristics of drying. In the past few years, some progress has been made in the study of the drying of bio-colloidal droplets building upon simple systems such as globular proteins in de-ionized water or saline buffer to whole blood – the most complex natural occurring bio-colloid solution [3-4]. Many researchers have investigated diseased blood samples such as thalassemia, anemia, etc., via this drying methodology. The goal there is to understand how and relate the final dried morphological patterns to the initial state of the bio-colloidal droplet with different diseased states [5]. The literature recently reported that the drying of a bio-colloidal droplet (at room temperature) has some of the constituent components pinned to the substrate and undergoes a flow during the initial drying process. As the drying continues and convective flow is firmly established, the fluid front recedes and deposits particles from the periphery towards the droplet's central region. The deposited particles during the fluid front movement forms a "coffee-ring" [6] near the edge of the original droplet and a layer throughout that finally results in crack patterns when drying is complete [3-4].

Lysozyme is a well-examined globular protein, and its solution in de-ionized water makes for the simplest bio-colloid as the lack of ions reduces protein-protein interactions. The drying process and the resulting convective flows are found to be affected by the atmospheric conditions, including temperature, relative humidity, substrate conditions, geometry, initial concentrations, and so on [1,7]. The non-uniform deposition of material after solvent evaporation along with the internal stresses that emerge upon the shrinkage of the film leads to a specific pattern of fractures. Whether or how the crack patterns emerge will depend on the initial state of the bio-colloidal solution [7]. However, a mound-like structure that forms in the central region is believed to be the leftover lysozyme particles carried along with the fluid front. This feature is believed to be a fingerprint for any aqueous lysozyme droplet [8,9]. Recently, the effects of the substrate temperature are observed on the protein-saline drying droplets. Different morphology is found in the various ranges of protein solutions and the substrate temperatures [10].

This experimental paper presents results that shed light on the (i) the self-assembling mechanism in the environment where both the initial concentration and temperature changes, (ii) whether it is possible to tune the final morphology by changing the dependent factors. This study is crucial for comprehending the protein-protein and protein-substrate interactions in different environmental conditions. For this, lysozyme's aqueous solution is prepared at different initial concentrations ($\emptyset$) ranging from 1 to 20 wt%. The drying evolution and the resulting morphology at various controlled substrate temperatures (T) of 25 to 55 °C are investigated using bright-field optical microscopy. The observations presented in this paper answers a few questions; can the formation of the mound-like feature in the lysozyme droplets be suppressed? If so, how? Is the coffee-ring effect always seen? What is the interplay between temperature and initial concentration? How is this interplay affecting (if any) the final dried morphological patterns?

2. **Materials and Experimental Methods**

Lysozyme is mostly found in the mucosal secretions of humans, such as tears, saliva, etc. The molecular mass of a lysozyme molecule is ~14.3 kDa with some apparent polydispersity. Each molecule has a roughly spherical shape of dimension $3.0 \times 3.0 \times 4.5$ nm$^3$. Its isoelectric point is 11.1, carrying a net positive charge under the present pH ~ 7 of this study [8, 9]. The lyophilized powder of hen-egg white lysozyme (HEWL, catalog no. L6876), a well-studied water-soluble globular protein, was used as obtained from Sigma-Aldrich Chemical Company (USA) without further processing. The bio-colloidal solution was prepared by dissolving the needed amounts of lysozyme in de-ionized water (Millipore, 18.2 MΩ.cm at 25 °C) to prepare the initial concentration (∅) of 20, 17, 13, 9, 5, 3, and 1 wt% at a pH of ~7. About 1 μL of the solution is pipetted on to a fresh, pre-cleaned, microscope coverslip (catalog no. 48366-045, VWR, USA) at room temperature of ~25 °C and relative humidity of ~50%. The hot stage mounted on a transmission mode bright-field optical microscope (Leitz Wetzlar, Germany) had been set to the desired observation temperature for a couple of hours prior to sample placement. These pipetted droplets were transferred to the stage within ~45 sec and equilibrated to the set temperature within about 10 sec. The temperature controller was set at different temperatures (T) of 25, 35, 45, and 55 °C with ±0.5 °C of uncertainty. The droplet images were captured under 5× magnification using an 8-bit digital camera (Amscope MU300) attached to the microscope every two seconds for about 30 minutes. The morphology of the final dried samples was captured after 24 hours. All the experiments were repeated three times at each temperature and concentration and exhibited high reproducibility. The captured images are then analyzed and quantified by various image processing techniques/macros in the ImageJ platform [11]. The pixels in the images were converted to an 8-bit grey scale and the pixel locations were converted into a mm-length scale using a length calibration slide. The fluid front radius was quantified five times on each image by drawing a line from the droplet center to a different point on its edge and recorded as a function of drying time [$\bar{r}(t)$]. The final similarly averaged radius of the droplets ($\bar{R}$), the averaged coffee-ring width ($\bar{w}$), the averaged crack spacing ($\bar{x}_c$), and their uncertainties are recorded for each T and ∅ experiment. The detailed procedure can be found in previous publications [2, 12].

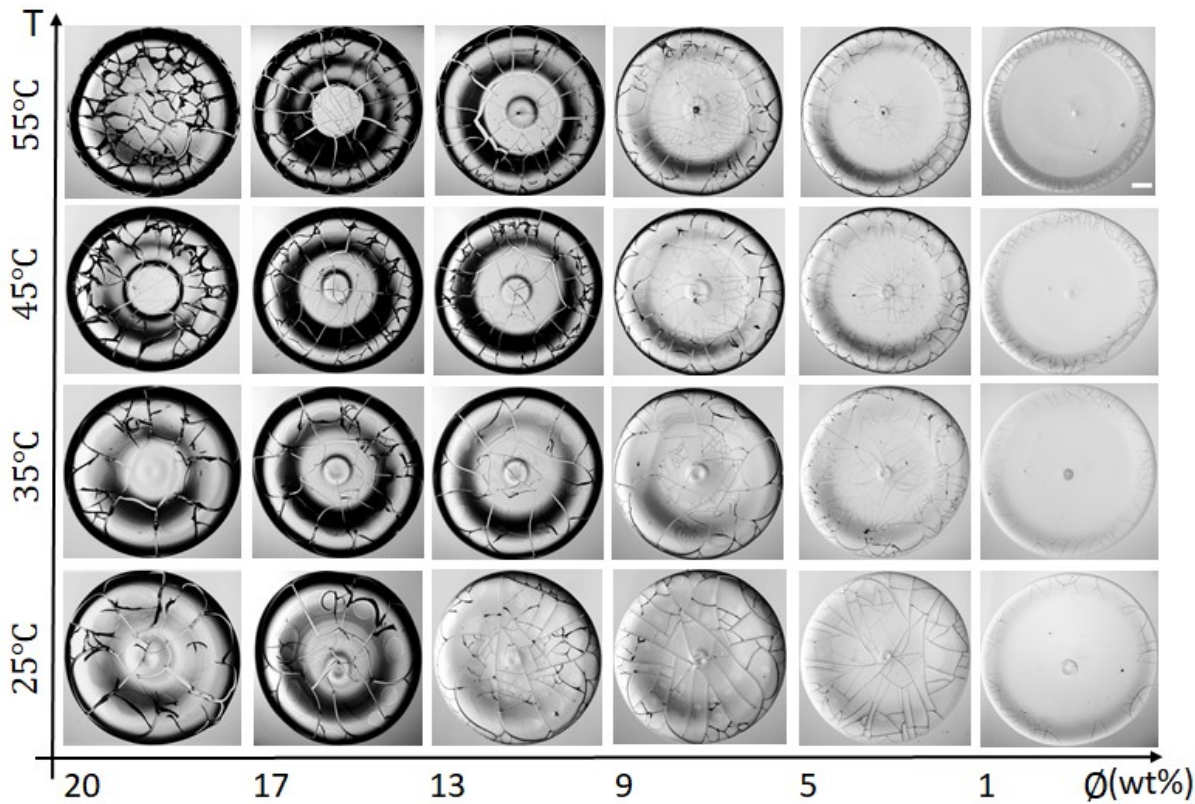

**FIG 1.** Morphology of dried lysozyme droplets captured within 24 hours at various initial concentrations (∅) diluted from 20 to 1 wt% at controlled substrate temperatures (T) of 25 to 55 °C. The white rectangle represents a scale bar of length 0.2 mm in the top-right panel 55 °C.

## 3. Results

Figure 1 shows the morphological patterns of lysozyme dried droplets at different initial concentrations (∅) from 20 to 1 wt% dried under controlled substrate temperatures (T) of 25 to 55 °C. The ∅ is divided into three regimes, ultra-concentrated (20 < ∅ ≤ 17 wt%), concentrated (17 < ∅ ≤ 9 wt%) and dilute (9 < ∅ ≤ 1 wt%). It should be noted that the morphological patterns shown here are somewhat different from our previous work [8], especially at the higher ∅, because a different substrate was used resulting in a different wetting angle, which plays an important role in determining these patterns.

A common trend observed is that the thick dark peripheral of the dried droplet in the ultra-concentrated regime becomes thin in the concentrated regime, and further diminishes in the diluted regime irrespective of temperature. The characteristic feature in all these samples at T = 25 °C is the presence of the central mound with a dimple (see the bottom panel of Fig. 1) [2]. However, the crack (fracture) patterns depend strongly on both T and ∅. In the ultra-concentrated regime, most of these radial cracks create large-sized domains. Some dark curved cracks are noticed near these radial cracks in this regime. In contrast, the random large and small-sized cracks are observed in the concentrated regime throughout the film. With more dilution, these random cracks are found only in the ring. The images at an elevated temperature of 35 °C are quite similar to those at 25 °C; however, the shadowy dark shade becomes sharper (mid panel of Fig. 1). An enlarged spot in the dimple is detected in various samples at elevated temperatures and ∅ of 9 to 1 wt%. Interestingly, no mound is observed at (T, ∅) = (45 and 55 °C, 20 wt%) and (55 °C, 17 wt%). Furthermore, the suppression of coffee-ring behavior and a unique crack pattern are seen at (T, ∅) = (55 °C, 20 wt%) [top panel of Fig. 1].

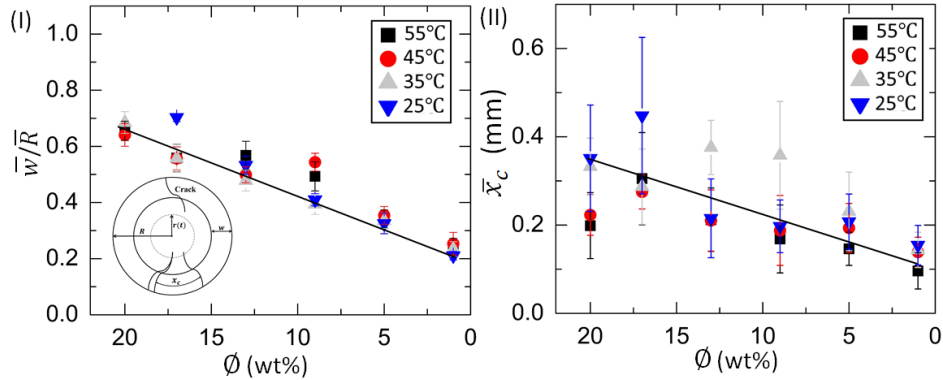

**FIG 2.** (I) shows the averaged ring width ($\bar{w}$) normalized with the averaged radius of the droplet ($\bar{R}$). (II) exhibits the averaged crack spacing ($\bar{x}_c$), both at various T and ∅. The error bars represent the standard deviation obtained from multiple measurements (n = 3). The inset of (I) displays a schematic droplet cartoon describing the radius of the fluid front ($r$) at an instant of time ($t$), the radius of the droplet ($R$), ring width ($w$) and crack spacing ($x_c$).

Figure 2 exhibits the concentration dependence of the normalized ring width ($\bar{w}/\bar{R}$) in (I) and the averaged spacing between the consecutive radial cracks ($\bar{x}_c$) in (II) at different controlled substrate temperature (T). It is found that both these parameters decrease with increasing ∅. However, the standard deviation of $\bar{w}/\bar{R}$ is not as large as $\bar{x}_c$ when plotted as a function of T and ∅. It is also to be noted that the quantification of the ring width is not possible at (T, ∅) = (55 °C, 20 wt%). It seems that the temperature does not play a strong role for the ring width shown in Fig. 2(I) as the data overlay each other closely. Similarly, the $\bar{x}_c$ behavior is also weakly dependent on temperature though being far more scattered with larger uncertainties for higher concentrations, which showed the largest separation between the two highest from the two lowest temperatures.

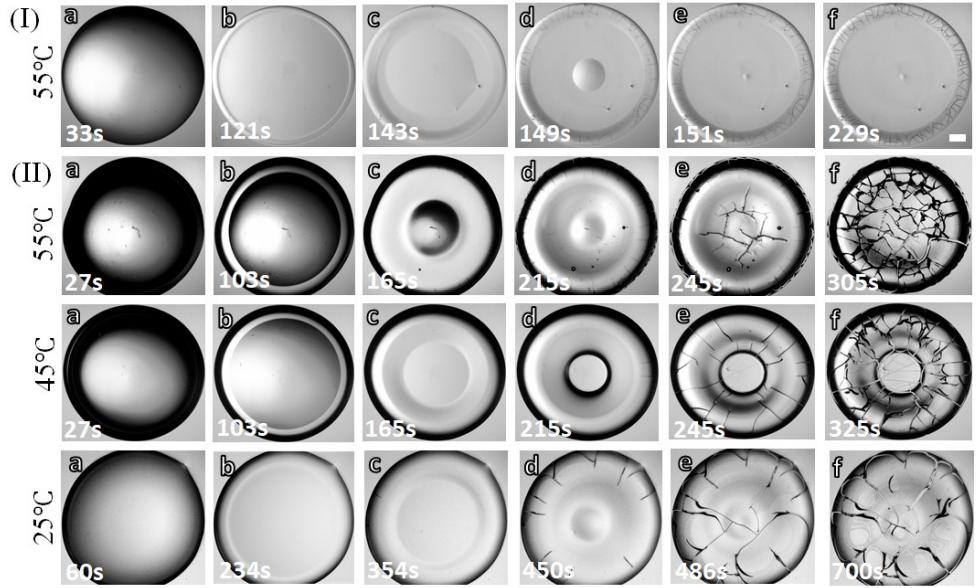

**FIG 3.** Drying evolution of lysozyme droplets: (I) a-f displays ∅ of 1 wt% drying under 55 °C. (II) a-f shows the ∅ of 20 wt% drying at 55, 45, and 25 °C. The timestamps are shown at the left-bottom of each image. The white rectangle represents a scale bar of length 0.2 mm in the top-right.

Figure 3(I-II) a-f describes the drying evolution of the selected lysozyme samples at (T, ∅) = (55 °C, 1 wt%) and (55, 45, and 25 °C, 20 wt%). Comparing Figs. 3(I) f and (II) f, the time required to complete the drying process at 55 and 45 °C (~290 sec) is less than half that at 25°C (~700 sec). For all the samples shown in Fig. 3(I-II) a-c, the same initial drying characteristics are seen, irrespective of T and ∅. Note that the time between the droplet deposition onto the coverslip and the first captured image is ~55 sec. These first images show a textural gradient, i.e., the dark texture near the periphery and bright near the center (see Fig. 3(I-II) a). However, the dark texture area is larger at 20 wt% than that of 1 wt%. As time progresses, the fluid front starts moving from the periphery to the center for all samples. A peripheral band is developed that is thickest for ∅ = 20 wt% (see Fig. 3(I-II) c). The front still moves inward when the cracks begin in the rim width regime (see Fig. 3(I-II) d-e). Only for the (T, ∅) = (55 °C, 1 wt%) sample and only in the ring area do small random cracks emerge. Interestingly, the 20 wt% sample shows a unique crack pattern for various T. Here, the cracks appear near the periphery and start propagating towards the center for (T, ∅) = (25 °C, 20 wt%) but reverses for the 20 wt% sample at T = 45 and 55 °C. Some of the curved cracks also develop to join these radial cracks at 25 °C. Almost no curved cracks are found at T = 45 °C. As T increases to 55 °C, the random various sized cracks originate near the center. Furthermore, the mound-like structure and the depression (dimple) within the mound appears are seen at (T, ∅) = (25 °C, 20 wt%) (a similar observation is reported in [8]). This mound then disappears at (T, ∅) = (45 °C, 20 wt%), and a spot is noticed in the mound at (T, ∅) = (55 °C, 1 wt%) [see Fig. 3(I-II) d-e]. The final morphology is displayed in Fig. 3(I-II) f.

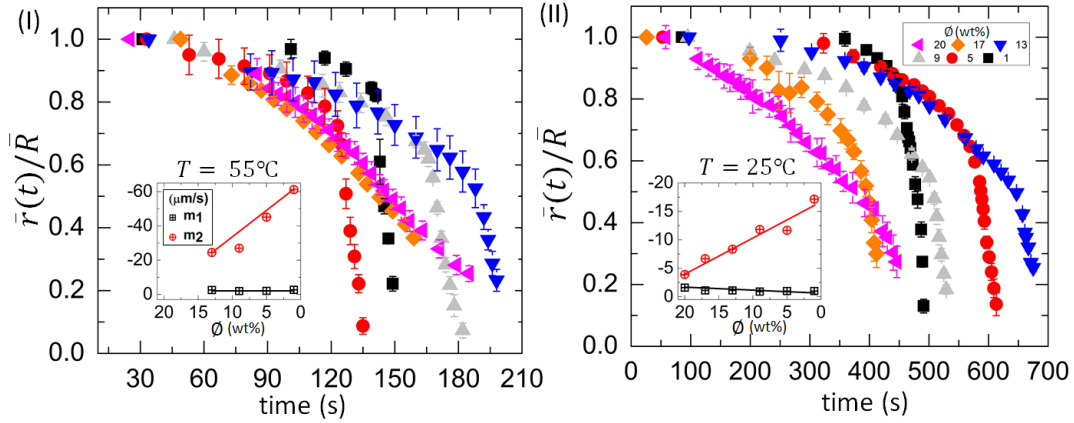

**FIG 4.** Time evolution of the averaged fluid front radius, $\bar{r}(t)$ normalized with the averaged radius of the droplet ($\bar{R}$) at an initial concentration (Ø) ranging from 20 to 1 wt% at 55 °C in (I) and 25 °C in (II). The error bars represent the standard deviation obtained from multiple measurements (n = 3). The inset shows the concentration dependence of the initial ($m_1$) and final ($m_2$) slope values at T = 55 °C in (I) and 25 °C in (II). These values are determined from the time evolution of the $\bar{r}(t)$.

Figure 4(I-II) shows the quantitative analysis of the fluid front movement during the drying evolution. The quantity, $\bar{r}(t)/\bar{R}$ is plotted as a function of time at 55 °C in Fig. 4(I) and 25 °C in Fig. 4(II). The fluid front moves slowly and linearly and then crosses over to a much faster movement in both dilute and concentrated regimes (reported in our earlier work [8]). Interestingly, the fluid front $\bar{r}(t)/\bar{R}$ moves almost linearly with nearly the same behavior over the entire time evolution for the ultra-concentrated samples at T = 55 °C, whereas there is a strong temperature dependence and two apparent rates for the lower concentration samples. The overall time until the disappearance of the droplet, for the fluid front to moves towards the center and vanish, is almost three times shorter at 55 °C compared to that at 25 °C. Two linear fits are done in the initial evolution and for the later time regime of the fluid front $\bar{r}(t)$ data to estimate the velocity of this front. The inset of Fig. 4(I-II) shows the concentration dependence of the initial ($m_1$) and later time ($m_2$) rate values at T = 25 and 55 °C. The averaged over all concentrations gives $m_1 = -1.12 \pm 0.25 \ \mu m \ s^{-1}$ at 25 °C and $-2.35 \pm 0.35 \ \mu m \ s^{-1}$ at 55 °C with very good regression $R^2$, being only weakly dependent on Ø. The negative sign indicates the reduction of the front radius with the progression of time. On the other hand, the extracted $m_2$ values increase linearly with increasing Ø. Note that the separate initial slow and later fast fluid front velocities are not apparent in the ultra-concentrated regime at 55 °C, where only a single rate is seen being $\bar{m} = -5.7 \ \mu m \ s^{-1}$ ($R^2 = 0.991$) for the 20 wt% sample and a similar $-5.1 \ \mu m \ s^{-1}$ ($R^2 = 0.987$) for the 17 wt% sample. Over the less concentrated samples, $m_2$ changes by $dm_2/d\emptyset = -3.17 \ \mu m \ s^{-1} wt\%^{-1}$ at 55 °C but only by $-0.64 \ \mu m \ s^{-1} wt\%^{-1}$ at 25 °C.

### 4. Discussions

The droplets are pipetted on the coverslip at 25 °C outside the microscope hot stage and then transferred to the hot stage well within ~45 sec. So, it is likely temperature fluctuations are present in these droplets, especially when the set temperature is ≥ 25 °C for the initial ~10 sec after closure. After the initial first minute of deposition and mounting, the subsequent imaging should be also affected by the temperature gradients driven by evaporative cooling and convective flow of the drying process. Since the images are taken from the top in transmission mode and captures the entire hemispherical-cap shape of the droplets, the uniformity of the drying conditions across the droplet is confirmed by the symmetrical fluid front radial movement and by the textural radial gradients (the dark region near the periphery and the bright region in the central region) seen in Fig. 3(I-II) a. This is because, due to droplet curvature, the evaporative mass loss is greatest near the three-phase contact line (interface between coverslip, droplet, and air) than the droplet top and drives the convective flow within the droplet during drying. The lysozyme protein globules are then deposited onto the coverslip at the fluid front by the convective flow, yielding the coffee ring distribution, as well as attraction to the glass throughout the droplet, yielding a film of protein inside this ring.

The key to understanding the observed drying and final dried patterns is the mass flow generated inside the droplet. The flow could be capillary, surface tension, and temperature-induced Marangoni flow. It is reported in the literature that if the Marangoni flow exceeds the outward radial flow, the ring-like behavior is suppressed [13]. However, all of our samples at each T show the ring-like behavior except (T, ∅) = (55 °C, 20 wt%) (see Figs. 1 and 3(I-II) a-f). Interestingly, we observe a dark peripheral band in each sample. This indicates that this suppression might not conclude the Marangoni flow's dominancy for every case. Therefore, we speculate that most of these lysozyme particles near the three-phase contact line are transported with the capillary flow. Energizing the system by increasing the T above the room temperature (here, up to 55 °C) might not dominate one flow over the other. The suppression at (T, ∅) = (55 °C, 20 wt%) is because there is not enough time to segregate (high T promotes the evaporation of water quickly) these large number of particles during the drying process. The droplets get pinned to the substrate. The thickness of this peripheral band is observed to be dependent on ∅. The thickest band is seen at ∅ = 20 wt%, and it becomes thin with the decreasing ∅ (see Figs. 1 and 3(I-II) b,c). This suggests that more material would be transported to the periphery for larger ∅. Simultaneously, the contact angle (angle formed by the droplet on the substrate) reduces. As soon as the contact angle minimizes to a large extent, the droplets show a gray texture. Meanwhile, the fluid front starts moving from the periphery towards the center. The front carries and deposits the material forming a ring-like behavior. The linear regime's slope values are small as the front carries more materials during the initial phase. As it carries less and less (after depositing along the line), the front speeds up, relating to the higher slope values. This also explains why ∅ = 1 wt% has the highest slope value in the non-linear regime during the fluid front movement [see Fig. 4(I-II)]. However, the absence of the non-linearity in the ultra-concentrated regime only at 55 °C indicates some interplay between T and ∅ exist in these lysozyme samples. It is more than just increasing the drying process rate, which is reported in our earlier work [14]. The leftover particles fall out of the solution giving rise to the mound-like structure in the central region. A depression (dimple) is also found, indicating the trapped water's evaporation within that mound. It is to be noted that all the lysozyme samples do not form this mound structure. As this structure can be turned off by supplying enough energy (increasing T) to the system, it should not be considered the lysozyme's fingerprint in general. The T was just enough that most of the particles at ∅ of 20 and 17 wt% are carried by the fluid front and there are no leftover particles by the end of its movement. This suppression is due to interplay of the T and ∅. This also indicates a phase diagram between T and ∅; however, two points are not enough, and need more samples to be investigated.

As a large amount of water is already evaporated, the mechanical stress develops due to the pinning effect, and the droplet is unable to shrink. The film cracks to relieve this stress (see Figs. 1 and 3(I-II) d,e). However, this crack propagation depends on when the available film stress exceeds the critical stress (popularly known as Griffith's criterion). For example, ∅ = 1 wt% does not show any cracks in the middle region (between the mound and the ring) [detailed in our previous paper, 8]. In contrast, the cracks are throughout the film in the concentrated samples. The morphological patterns are different in the ultra-concentrated samples since the thickness gradient throughout the film differs at elevated T. Though this study provides clues for the interplay of the T and ∅ affecting the final morphology, it would be interesting to study these samples under infrared microscopy. This will allow us to better understand the physics by observing the isotherms on the droplet-air interface, the substrate's surface, and mapping this temperature distribution at each timestamp during the drying process.

## 5. Conclusion

This paper concludes that the underlying physical mechanism is vital to understanding the emergent morphological patterns in any bio-colloidal drying droplets. This study shows that the mound-like structure is not the lysozyme's characteristics in all drying conditions. It can be turned off by energizing the system, i.e., increasing the substrate temperature. We understand that the high range of initial lysozyme concentration and the sample preparation in de-ionized water is way too far from an *in-vivo* environment and might not be appreciable in the context of biology. However, we hope that the rich physics involved in this type of bio-colloidal systems will encourage soft-matter physicists to do detailed investigations to understand the interplay between concentration and temperature on the protein-protein and protein-substrate interactions.